\documentclass[showpacs,showkeys,preprintnumbers,amsmath,amssymb,latexsym]{revtex4}

\usepackage{dcolumn}

\def\rmd{{\rm d}} 

\def\beq{\begin{equation}}
\def\eeq{\end{equation}}

\catcode`@=11  
\def\meqalign#1{\null\,\vcenter{\openup\jot\m@th
\ialign{\strut\hfil$\displaystyle{##}$&&$\displaystyle{{}##}$\hfil
\crcr#1\crcr}}\,}
\catcode`@=12  

\def\pmb#1{\setbox0=\hbox{$#1$}%
  \kern-.025em\copy0\kern-\wd0
  \kern.05em\copy0\kern-\wd0
  \kern-.025em\raise.0433em\box0}
 
\begin{document}

\preprint{Version: March, 30  2004}

\title{Vacuum C-Metric and the Gravitational Stark Effect}

\author{D. Bini}
\email{binid@icra.it}
\affiliation{Istituto per le Applicazioni del Calcolo ``M. Picone'', CNR, I-00161 Rome, Italy and\\ 
International Center for Relativistic Astrophysics - I.C.R.A.\\
University of Rome ``La Sapienza'', I-00185 Rome, Italy}

\author{C. Cherubini}
\email{cherubini@icra.it}
\affiliation{Faculty of Engineering, University Campus Bio-Medico of Rome,  I-00155 Rome, Italy\\
Institute of Cosmology and Gravitation, University of Portsmouth, Portsmouth PO1 2EG, England, UK\\
Dipartimento di Fisica ``E.R. Caianiello'', Universit\`a di Salerno, I-84081, Italy and\\
International Center for Relativistic Astrophysics - I.C.R.A.\\
University of Rome ``La Sapienza'', I-00185 Rome, Italy}

\author{B. Mashhoon}
\email{MashhoonB@missouri.edu}
\affiliation{Department of Physics and Astronomy,
University of Missouri-Columbia, Columbia,
Missouri 65211, USA}

\date{\today}

\begin{abstract}
We study the vacuum C-metric and its physical interpretation in terms of the exterior spacetime of a uniformly accelerating spherically - symmetric gravitational source. Wave phenomena on the linearized C-metric background are investigated. It is shown that the scalar perturbations of the linearized C-metric correspond to the gravitational Stark effect. This effect is studied in connection with the Pioneer anomaly. 
\end{abstract}



\maketitle

\section{Introduction}

In testing general relativity within the solar system, the post-Newtonian framework is generally employed 
(see, e.g., \cite{sha}). The post-Newtonian formalism is essentially based on an underlying inertial system of coordinates \cite{will}, which for solar-system studies can be identified with the barycentric reference frame. This assumes that the center of mass of the Sun follows a geodesic. In this paper, we raise the possibility that the center of mass of the Sun may be undergoing translational acceleration due to nongravitational forces resulting perhaps from anisotropic solar emission. Within the framework of general relativity, a free rotating extended body does not follow a geodesic according to the Mathisson-Papapetrou-Dixon equations; however, the deviation from a geodesic for the Sun or the planets would be negligibly small for our present considerations. We therefore neglect the gravitational effects of the rotation of the Sun in this work and concentrate instead on the possibility that the center of mass of the Sun has a small but non-negligible translational acceleration of unknown origin. Thus the barycentric reference frame would then be an accelerated frame of reference. All planets and satellites would be affected due to the presence of inertial forces arising from the noninertial character of the reference system. The effect of such inertial forces on the planets could be negligible at present in comparison with other forces; regarding satellites however, it may be the source of the Pioneer anomaly as discussed later in this paper.

The exterior gravitational field of a uniformly accelerated spherically symmetric gravitational source is given by the vacuum C-metric. This metric is discussed in section II. We are particularly interested in the motion of particles and waves in this field. To this end, we first express the C-metric in appropriate coordinates as a nonlinear superposition of the Schwarzschild and Rindler metrics. Then in section III we linearize the C-metric and show that the propagation of particles and waves in the linearized C-metric corresponds to the gravitational Stark effect. That is, a particle in this field is subject to the gravitational attraction of the central source as well as the uniform inertial force; this is the gravitoelectric analog of an electron in the combined Coulomb and an external uniform electric field, as in the Stark effect. The consequences of this effect for interferometry in the gravitational field of an accelerated object are pointed out in section IV. The Pioneer anomaly is discussed in section V. Section VI contains a brief discussion of our results.

\section{Vacuum C-Metric}

The vacuum C-metric was first discovered by Levi-Civita \cite{LC} in 1918 within a class of degenerate static vacuum metrics.
However, over the years it has been rediscovered many times: by Newman and Tamburino \cite{newtam} in 1961, by Robinson and Trautman \cite{robtra} in 1961 and again by Ehlers and Kundt \cite{ehlkun} ---who called it the C-metric---
in 1962. The charged C-metric has been studied in detail by Kinnersley and Walker \cite{kin69,kinwal}. In general the spacetime represented by the C-metric contains one or, via an extension, two uniformly accelerated particles as explained in \cite{kinwal,bon83}.
A description 
of the geometric properties of various extensions of the C-metric as well as a more complete list of references is contained in \cite{ES}.
The main property of the C-metric is the existence of two hypersurface-orthogonal  Killing vectors, one of which is
timelike (showing the static property of the metric) in the spacetime region of interest in this work.
The C-metric is often written in the form \cite{kin69,kinwal}
\begin{equation}
\label{met_txyz}
\rmd s^2=\frac{1}{A^2(\tilde x+\tilde  y)^2}[(\tilde F \rmd t^2 - \tilde F^{-1}\rmd \tilde y{}^2) -(\tilde G^{-1} \rmd \tilde x{}^2 + \tilde G\rmd \tilde z{}^2)],
\end{equation}
where
\begin{equation}
\tilde F(\tilde y):= -1+\tilde y{}^2-2mA\tilde y{}^3,\qquad  \tilde G(\tilde  x):= 1-\tilde x{}^2-2mA\tilde x{}^3, \qquad \tilde G(\tilde x)=-\tilde F(-\tilde x).
\end{equation}
These coordinates are adapted to the hypersurface-orthogonal Killing vector $\kappa=\partial_t$, the spacelike Killing vector
$\partial_{\tilde z}$ and $\partial_{\tilde x}$, which is aligned along the non-degenerate eigenvector of the hypersurface Ricci tensor.
The constants $m\ge 0$ and $A\ge 0$ denote the mass and acceleration of the source, respectively. 
Unless specified otherwise, we choose units such that the gravitational constant and the speed of light in vacuum are unity.
Moreover, we assume that the C-metric has signature -2; to preserve this signature, we must have $\tilde G>0$. We assume further that $\tilde F>0$; it turns out that the physical region of interest in this case corresponds to $mA<1/(3\sqrt{3})$ \cite{Farh,pavda,podol} .
The C-metric is of Petrov type D and belongs to the Weyl class of solutions of the Einstein equations \cite{ES}.

It is useful to introduce the retarded coordinate $u$, the radial coordinate $r$ and the azimuthal coordinate $\phi$:
\begin{equation}
u=\frac1A [t+\int^{\tilde y} \tilde F^{-1}\rmd \tilde y], \qquad r=\frac{1}{A(\tilde  x+\tilde  y)}, \qquad \phi= \tilde z ,
\end{equation}
so that the metric can be cast in the form
\begin{equation}
\label{cmetEF}
\rmd s^2= \tilde H \rmd u^2 + 2 \rmd u \rmd r + 2A r^2 \rmd u \rmd \tilde x -\frac{r^2}{\tilde G} \rmd \tilde x{}^2 -r^2\tilde G \rmd \phi^2 , 
\end{equation}
where
\begin{equation}
\tilde H(r,\tilde x)=1-\frac{2m}{r}-A^2r^2 (1-\tilde x{}^2-2mA\tilde x{}^3)-Ar(2\tilde x+6mA\tilde x{}^2)+6mA\tilde  x .
\end{equation}
The norm of the hypersurface-orthogonal Killing vector $\kappa$ is determined by $\tilde H$, 
\begin{equation}
\kappa_\alpha \kappa^\alpha = r^2 \tilde F=\frac{\tilde H}{ A^2} ,
\end{equation}
so that this Killing vector is timelike for $\tilde H>0$.

In this section, we deal separately with the two limiting cases of $m = 0$ and $A = 0$; therefore, we find it
 convenient to work with the $\{u,r,\theta,\phi \}$ coordinate system, where 
$(r,\theta , \phi)$ are spherical polar coordinates with  $\tilde x=\cos \theta$. Thus the C-metric takes the form
\begin{equation}
\label{cmetu}
\rmd s^2= H \rmd u^2 + 2 \rmd u \rmd r - 2A r^2 \sin \theta \rmd u \rmd \theta -\frac{r^2\sin^2\theta }{G} \rmd \theta^2 -r^2 G \rmd \phi^2  ,
\end{equation} 
where $G$ and $H$ are given by
\begin{eqnarray}
G(\theta)&=& \sin^2 \theta -2mA \cos^3\theta\, , \nonumber \\
H(r,\theta)&=&1-\frac{2m}{r}-A^2r^2 (\sin^2\theta-2mA\cos^3\theta)-2Ar\cos\theta(1+3mA\cos\theta)+6mA\cos\theta .
\end{eqnarray}
Bonnor \cite{Bonnor2} has discussed how the present form of the C-metric can be cast in Bondi's form.

As already demonstrated in \cite{kin69,kinwal}, the metric (\ref{cmetu}) can be seen to be a nonlinear superposition of two metrics, one  associated with a Schwarzschild black hole (case $A=0$) and the other corresponding to a uniformly accelerating particle (case $m=0$). To illustrate this point we proceed as follows.
Let us first consider a background Minkowski spacetime with Cartesian (inertial) coordinates $x^\mu=\{t,x,y,z \}$ and imagine a {\it point mass} $m$ ---assumed to be only a {\it test} particle at first--- accelerating along the negative $z$-axis with uniform acceleration $A$.
The worldline of the test particle, parametrized with its proper time $\tau$, is given by
\begin{equation}
t= \frac1A \sinh A\tau , \quad x=0, \quad y=0, \quad z=z_0- \frac1A (-1+\cosh A\tau).
\end{equation}
It is convenient to assume $z_0=-1/A$, so that  the worldline of $m$ can be expressed as
\begin{equation}
x^\mu_m= \frac1A (\sinh A\tau , 0,0, -\cosh A\tau ).
\end{equation}
Let us now set up a Fermi frame (i.e. a nonrotating Fermi-Walker transported tetrad) along this worldline; that is,  
\begin{eqnarray}
\lambda_{(0)}&=& \cosh A\tau \partial_t - \sinh A\tau \partial_z, \nonumber \\
\lambda_{(1)}&=& \partial_x , \nonumber \\
\lambda_{(2)}&=& \partial_y , \nonumber \\
\lambda_{(3)}&=& - \sinh A\tau \partial_t +\cosh A\tau \partial_z .
\end{eqnarray}
The Fermi coordinates $\{T,X,Y,Z\}$ are defined to be such that
\begin{equation}
x^\mu -x^\mu_m=X^i \lambda_{(i)}^\mu, \qquad \tau=T.
\end{equation}
Therefore the map $\{t,x,y,z\}\rightarrow \{T,X,Y,Z\}$ is given by
\begin{eqnarray}\label{COORDYS}
t&=& (\frac1A -Z)\sinh AT, \nonumber \\
x&=&X , \nonumber \\
y&=&Y , \nonumber \\
z&=& -  (\frac1A -Z)\cosh AT.
\end{eqnarray}
With this coordinate transformation, the line element $\rmd s^2=\rmd t^2-\rmd x^2-\rmd y^2-\rmd z^2$ 
becomes the Rindler metric \cite{R,R1}
\begin{equation}
\label{I}
\rmd s^2=(1-AZ)^2\rmd T^2-\rmd X^2-\rmd Y^2-\rmd Z^2.
\end{equation}
The coordinates in (\ref{I}) are admissible for $T\in (-\infty , +\infty)$, $X\in (-\infty , +\infty)$, $Y\in (-\infty , +\infty)$ and $Z\in (-\infty , 1/A)$.
In the Rindler spacetime, the hypersurface-orthogonal Killing vector $\partial_T$ is timelike, but becomes null on the horizon $Z=1/A$. The coordinate system breaks down beyond this limit and is inadmissible. Actually $Z=1/A$ corresponds to a null cone in  $\{t,x,y,z\}$ coordinates, 
since  $z^2-t^2=(Z-1/A)^2$.

Let us now suppose  that the point mass $m$ is {\it not} a test particle; that is, 
its gravitational field cannot be neglected.
In the Fermi system the point particle is at the origin of coordinates $(T,0,0,0)$; 
therefore, by itself, its metric should be the Schwarzschild solution given by
\begin{equation}
\label{II}
\rmd s^2 = \left(1-\frac{2m}{r}\right)\rmd T^2 -\left(1-\frac{2m}{r}\right)^{-1}\rmd r^2 - r^2 \rmd \Omega^2,
\end{equation}
where $\rmd \Omega^2=\rmd \theta^2 +\sin^2 \theta \rmd \phi ^2$ and
\begin{equation}
X=r \sin \theta \cos \phi, \quad Y=r \sin \theta \sin \phi, \quad Z=r \cos \theta. 
\end{equation}
We will show that the uncharged C-metric is a nonlinear superposition of (\ref{I}) and (\ref{II}).
To this end, 
we consider the C-metric written in the form (\ref{cmetu}).

Case 1. Let $A=0$. Then (\ref{cmetu}) becomes
\begin{equation}
\label{IIu}
\rmd s^2 = \left(1-\frac{2m}{r}\right)\rmd u^2 +2 \rmd u\rmd r - r^2 \rmd \Omega^2 .
\end{equation}
Next, we introduce the coordinate transformation $u=-T-r^*$, where $r^*$ is the tortoise coordinate
\begin{equation}
r^*=r+2m \ln \left( \frac{r}{2m}-1\right), \qquad \frac{\rmd r^*}{\rmd r}=\left( 1-\frac{2m}{r}\right)^{-1} .
\end{equation}
With $\rmd u = -\rmd T- \rmd r /(1-\frac{2m}{r})$, $\rmd s^2$ takes the form (\ref{II}), i.e. it coincides with the Schwarzschild solution. Of course in this case one has a horizon at $r=2m$.
The hypersurface-orthogonal Killing vector $\partial_T$ is timelike for $r>2m$, null on the horizon $r=2m$ and spacelike in the interior $r<2m$.

Case 2. Let $m=0$. Then the metric  (\ref{cmetu}) becomes
\begin{equation}
\label{formeq0}
\rmd s^2= (1-2Ar\cos \theta -A^2r^2 \sin^2\theta )\rmd u^2 +2 \rmd u \rmd r -2Ar^2 \sin \theta \rmd u \rmd \theta  -r^2 \rmd \theta^2 -r^2 \sin^2 \theta \rmd \phi^2.
\end{equation}
By using the coordinate transformation
\begin{eqnarray}
\label{uXYZ}
T&=&-u+\frac{1}{2A}\ln \left[\frac{1-Ar (1+\cos \theta)}{1+Ar (1-\cos \theta)}\right], \nonumber \\
X&=&r \sin \theta \cos \phi, \nonumber \\
Y&=&r \sin \theta \sin \phi, \nonumber \\
Z&=&\frac1A -\sqrt{\left(\frac1A-r\cos \theta\right)^2-r^2},
\end{eqnarray}
the metric (\ref{I}) takes the form (\ref{formeq0}). To see how this comes about, let us  start with (\ref{I}) and introduce new coordinates $T$ and $Z$ according to
\begin{eqnarray}
-\left(\frac1A-Z \right)\sinh AT &=& \left(\frac1A- r\cos \theta \right)\sinh Au + r \cosh Au , \nonumber \\
\left(\frac1A-Z \right)\cosh AT &=& \left(\frac1A- r\cos \theta \right)\cosh Au + r \sinh Au . 
\end{eqnarray}
By adding and subtracting these equations one gets
\begin{eqnarray}
\left(\frac1A- Z \right)&=& \left[\frac1A+ r (1-\cos \theta) \right]e^{A(u+T)}, \nonumber \\
\left(\frac1A- Z \right)&=& \left[\frac1A- r (1+\cos \theta) \right]e^{-A(u+T)}. 
\end{eqnarray}
Now multiply and divide these equations to find
\begin{eqnarray}
\left(\frac1A- Z \right)^2= \left(\frac1A- r\cos \theta  \right)^2-r^2, \nonumber \\
\frac{1+A r (1-\cos \theta)}{1-A r (1+\cos \theta)}=e^{-2A(u+T)},  
\end{eqnarray}
from which the expressions for $Z$ and $T$ in (\ref{uXYZ}) follow.
We note that in the relation for $Z$ in (\ref{uXYZ}) the sign has been chosen such that $Z\to r\cos\theta$ as $A\to 0$.
The coordinates $T$ and $Z$ are admissible for $r_-<r<r_+$, where
\begin{equation}
r_\pm = \pm \frac{1}{A(1\pm \cos \theta)}.
\end{equation}
If we interpret $r$ as the radial coordinate, then 
\begin{equation}
\label{bou}
0\le r<r_+=\frac{1}{A(1+\cos \theta)}.
\end{equation}
The boundary region corresponds to the horizon in $(r, \theta)$ coordinates.
It is natural to assume that in the nonlinear superposition of these two cases that results in the vacuum C-metric, the horizons at $r=2m$ and $r=r_+$ would be appropriately modified. This can be deduced explicitly from the form of $H$.

It is useful to introduce an acceleration lengthscale based on $A>0$ given by
\begin{equation}
L_A= \frac{1}{3\sqrt{3}A}. 
\end{equation}
It turns out that the modification of the horizons is related to the ratio
of $m$ and $L_A$.
The event horizons of the vacuum C-metric are Killing horizons given by $H=0$ \cite{kinwal}.
The solution of $H=0$ can be written as 
\begin{equation}
r^{-1}=A (\cos \theta + W^{-1}),
\end{equation}
where $W$ is a solution of $W^3-W+2mA=0$. There are three cases depending on whether $m$ is less than, equal or greater than $L_A$. We have assumed at the outset that $m<L_A$; therefore, we expect that the two individual horizons, i.e. the inner one at $r=2m$ and the outer one at $r=r_+$ will be somewhat modified. In fact let
\begin{equation}
\frac{1}{\sqrt{3}}\left( -\frac{m}{L_A}+i\sqrt{1-\frac{m^2}{L_A^2}}\right) =\hat U+i \hat V,
\end{equation}
then there are three real solutions for $W$ given by
$W=2\hat U$, which results in $r=2m$ for $A\to 0$, $W=-\hat U-\sqrt{3}\hat V$, which results in $r_+^{-1}=A(1+\cos\theta)$ for $m\to 0$,
and  $W=-\hat U+\sqrt{3}\hat V$, which results in $r_-^{-1}=A(\cos\theta -1 )$ for $m\to 0$ and is therefore unacceptable.

\section{Gravitational Stark effect}

It is interesting to reduce the C-metric to linear form in $m$ and $A$ by neglecting $m^2$, $mA$, $A^2$ and higher-order terms.
It follows from (\ref{uXYZ}) in the $m=0$ limit that
\begin{eqnarray}
T&=& -u-r-Ar^2 \cos\theta +O(A^2), \nonumber \\
Z&=& r\cos \theta +\frac12 A r^2+O(A^2). 
\end{eqnarray}
Therefore, in metric (\ref{cmetu}) let us consider the coordinate transformation $\{u,r,\theta, \phi\}\rightarrow \{T,X,Y,Z\}$, 
where 
\begin{eqnarray}
\label{approxuXYZ}
T&=&-u- \left[r+2m \ln \left(\frac{r}{2m}-1\right)\right]-Ar^2\cos \theta, \nonumber \\
X&=&r \sin \theta \cos \phi, \nonumber \\
Y&=&r \sin \theta \sin \phi, \nonumber \\
Z&=& r\cos \theta +\frac12 Ar^2.
\end{eqnarray}

This is a \lq\lq linear superposition" of the transformations used in Case 1 and Case 2 of the previous section. The metric (\ref{cmetu}) under the transformation (\ref{approxuXYZ}) takes the form
\begin{equation}
\label{newmet}
\rmd s^2= (1-\frac{2m}{R}-2AZ)\rmd T^2 -\frac{2m}{R^3} (X\rmd X+Y\rmd Y+Z\rmd Z )^2 -\rmd X^2 -\rmd Y^2 -\rmd Z^2 ,
\end{equation}
where $R=\sqrt{X^2+Y^2+Z^2}$ and we have neglected $m^2$, $mA$, $A^2$ and higher - order terms.
Next, introduce polar coordinates $\Theta$ and $\Phi$ such that
\begin{eqnarray}
X=R\sin \Theta \cos \Phi,\quad Y=R \sin \Theta \sin \Phi,\quad
Z=R\cos \Theta .
\end{eqnarray}
With respect to these, our linearized metric becomes
\begin{equation}
\label{linearizzata}
\rmd s^2= \left(1-\frac{2m}{R}-2AR\cos \Theta \right)\rmd T^2 -\left(1+\frac{2m}{R}\right) \rmd R^2 -R^2 (\rmd \Theta^2 +\sin^2 \Theta \rmd \Phi^2) ,
\end{equation}
which is a linear superposition of (\ref{I}) and (\ref{II}). It is useful to note here the transformation $\{u,r,\theta,\phi\}\rightarrow \{T,R,\Theta, \Phi\}$ given by
\begin{eqnarray}
u & =& -T  -R-2m\ln (\frac{R}{2m}-1)-\frac12 AR^2 \cos\Theta ,\nonumber \\
r&=& R -\frac12 AR^2 \cos\Theta ,\nonumber \\
\theta&=& \Theta + \frac12 AR \sin\Theta ,\nonumber \\
\phi&=&\Phi
\end{eqnarray}
that takes  metric (\ref{cmetu}) directly to the form (\ref{linearizzata}) neglecting terms of order $m^2$, $mA$, $A^2$, etc.
Finally, introducing the isotropic radial coordinate $\rho$,
\begin{equation}
R=\left(1+\frac{m}{2\rho}\right)^2\rho=\rho +m +\frac{m^2}{4\rho}\simeq \rho+m\, ,
\end{equation}
we get the linear metric in standard form
\begin{equation}
\label{iso}
\rmd s^2= \left(1-\frac{2m}{\rho}-2A\hat Z\right)\rmd T^2 -\left(1+\frac{2m}{\rho}\right) (\rmd \hat X{}^2 +\rmd \hat Y{}^2+\rmd \hat Z{}^2), 
\end{equation}
where
\begin{equation}
 \hat X= \rho \sin \Theta \cos \Phi , \quad \hat Y= \rho \sin \Theta \sin \Phi, \quad \hat Z= \rho \cos \Theta .
\end{equation}
The geodesic motion of a test particle in this gravitational field, which  is characterized by a gravitoelectric potential $m/\rho+A \hat Z$, is quite similar to the classical motion of the electron in the Stark effect. Moreover, the wave mechanics of the Stark effect is also reflected in the behavior of the perturbations of (\ref{iso}) by a classical massless radiation field.
Consider the massless scalar field equation
\begin{equation}
\label{lapla}
\nabla^\mu \partial_\mu \chi =0
\end{equation}
on the background spacetime given by the metric (\ref{iso}). 
To first order in $m$ and $A$, $\chi$ can be separated by introducing parabolic coordinates  in analogy with the Stark effect, which is
the shift in the energy levels of an atom in an external electric field corresponding to the eigenvalues of a Schr\"odinger equation with
a Coulomb potential $-k/r$ plus the potential due to a constant electric field $\mathbf{E}=E \, \hat{\mathbf{z}}$, i.e. $-k/r + eE z$, where  $-e$ is the charge of the electron. 
In this gravitoelectromagnetic counterpart of the Stark effect, we set
\begin{equation}
\hat X = \sqrt{\xi \eta} \cos \psi, \quad \hat Y = \sqrt{\xi \eta} \sin \psi, \quad \hat Z =\frac12 (\xi-\eta), 
\end{equation}
and assume that
\begin{equation}
\label{Psi}
\chi(T,\xi,\eta,\psi)= e^{-i \omega T}\,  e^{i \nu \psi} U(\xi) V(\eta),
\end{equation}
where $\xi\ge 0$, $\eta\ge 0$, $\psi$ takes values from $0$ to $2\pi$, $\omega$ is a constant and $\nu$ is an integer.

It follows from (\ref{lapla}) that
\begin{eqnarray}
\label{eqUV}
&& U_{\xi\xi}+\frac{1}{\xi}\left( 1-\frac12 A\xi \right)U_\xi+\left[\frac{\omega^2}{4}(1+\xi A) + \frac{1}{\xi}(m\omega^2 -C)-\frac{\nu^2}{4\xi^2}\right]U=0 , \nonumber \\
&& V_{\eta\eta}+\frac{1}{\eta}\left( 1+\frac12 A\eta \right)V_\eta+\left[\frac{\omega^2}{4}(1-\eta A) + \frac{1}{\eta}(m\omega^2 +C)-\frac{\nu^2}{4\eta^2}\right]V=0,
\end{eqnarray}
where $C$ is the separation constant and $U_\xi=\rmd U /\rmd \xi$, etc.
Note that the second equation for  $V(\eta)$ can be obtained from the first one for $U(\xi)$ by replacing $A\to -A$ and $C\to -C$.
Introducing a new constant $\beta$ by
\begin{equation}
C=\frac12\left(\beta -\frac{A}{2}\right)
\end{equation}
and rescaling $U$ and $V$,
\begin{equation}
U(\xi) =\left(1+\frac{A\xi}{4}\right)a(\xi) , \qquad V(\eta)=\left(1-\frac{A\eta}{4}\right)b(\eta) ,
\end{equation}
eqs. (\ref{eqUV}) become
\begin{eqnarray}
\label{eqab}
&& \frac{d}{d\xi} \left(\xi \frac{da}{d\xi}\right)+ \left[\frac{\omega^2 \xi}{4}-\frac{\nu^2}{4\xi}+\frac{A\omega^2}{4}\xi^2 +\left(m\omega^2+\frac{\beta}{2}\right)\right]a=0,\nonumber \\
&& \frac{d}{d\eta} \left(\eta \frac{db}{d\eta}\right)+ \left[\frac{\omega^2 \eta}{4}-\frac{\nu^2}{4\eta}-\frac{A\omega^2}{4}\eta^2 +\left(m\omega^2-\frac{\beta}{2}\right)\right]b=0.
\end{eqnarray}

These equations can be put in exact correspondence with the  Schr\"odinger equation for the hydrogen atom in a constant electric field that results in the Stark effect \cite{LLMQ}
\begin{equation}
i\partial_t \Psi =(\mathcal{H}_0+\mathcal{H}_1)\Psi, 
\end{equation}
where we set $\hbar=1$ in this section and $\Psi$ is given by $\Psi(t,\xi,\eta,\psi)= e^{-i E_0 t}\,  e^{i \nu \psi} a(\xi) b(\eta)$,
\begin{equation}
\mathcal{H}_0 = -\frac{1}{2M} \nabla^2 -\frac{k}{r}, \qquad \mathcal{H}_1=eEz= \frac{\mathcal{E}}{M}z\ 
\end{equation}
and the standard notation for $r$, related to the Cartesian coordinates by $r=(x^2+y^2+z^2)^{1/2}$, has been used.
Here $\mathcal{E}$ is a new perturbation parameter defined by $\mathcal{E}=eME$. For the hydrogen atom $M$ is the reduced mass and $k=e^2$.
Passing to parabolic coordinates
\begin{equation}
x = \sqrt{\xi \eta} \cos \psi, \quad y = \sqrt{\xi \eta} \sin \psi, \quad z =\frac12 (\xi-\eta), 
\end{equation}
one has $r=(\xi+\eta)/2$ and
\begin{equation}
\nabla^2= \frac{4}{\xi+\eta}\left[\frac{\partial}{\partial\xi} \left(\xi \frac{\partial}{\partial\xi}\right)+
\frac{\partial}{\partial\eta} \left(\eta \frac{\partial}{\partial\eta}\right)\right]+\frac{1}{\xi\eta}\frac{\partial^2}{\partial\psi^2}.
\end{equation}
The equations for $a$ and $b$ then become
\begin{eqnarray}
\label{eqabclass}
&& \frac{d}{d\xi} \left(\xi \frac{da}{d\xi}\right)+ \left[\frac{ME_0 \xi}{2}-\frac{\nu^2}{4\xi}-\frac{\mathcal{E}}{4}\xi^2 +\frac{M}{2}(k+\tilde\beta)\right]a=0,\nonumber \\
&& \frac{d}{d\eta} \left(\eta \frac{db}{d\eta}\right)+ \left[ \frac{ME_0 \eta}{2}-\frac{\nu^2}{4\eta}+\frac{\mathcal{E}}{4}\eta^2 +\frac{M}{2}(k-\tilde\beta)\right]b=0, 
\end{eqnarray}
where $\tilde\beta$ is the separation constant.
The correspondence of these equations with (\ref{eqUV}) is {\it exact} once
\begin{equation}
ME_0 \to \frac{\omega^2}{2},\qquad \mathcal{E} \to -A\omega^2, \qquad k M \to 2m\omega^2, \qquad M\tilde\beta\to\beta.
\end{equation}

The solution of  equations (\ref{eqabclass}) is discussed in detail in standard treatments of nonrelativistic quantum mechanics (see, e.g. \cite{LLMQ}).
In the absence of the perturbation (i.e. $\mathcal{E}=0$), it is well known that the solutions of (\ref{eqabclass}) can be expressed in terms of the hypergeometric functions for the discrete ($E_0<0$) as well as the continuum ($E_0>0$) states of the hydrogen atom. Equations (\ref{eqabclass}) cannot be solved exactly; however, the Stark shift can be evaluated in perturbation theory for a sufficiently small external electric field.

Let us note here again  the close formal correspondence between the  quantum theory of the Stark effect in hydrogen and the theory of a classical massless scalar field on the linearized C-metric background. 
To extend this result to massless fields with nonzero spin, we encounter difficulties. In fact, using the Newman-Penrose formalism, it is possible to separate the equations in a particular set of coordinates \cite{prestidge}; on the other hand, the field equations are not separable in the physically transparent coordinate system of our linearized C-metric.

For many laboratory applications, the potential associated with the gravitational Stark effect can be written as $m/\rho+A\rho \cos \Theta$ with
$\rho=\rho_\oplus+\zeta$, where $\rho_\oplus$ is the average radius of the Earth and $\zeta$ is the local vertical coordinate in the laboratory. Using the local acceleration  of gravity, $g=m/\rho_\oplus^2$, the effective {\it Newtonian} gravitational potential is then $-m/\rho_\oplus +g\zeta -A(\rho_\oplus+\zeta)\cos \Theta$; some of the applications of this potential are discussed in the next section. 

\section{Acceleration-induced phase shift}

It follows from the results of the previous section that wave phenomena in the exterior spacetime represented by (\ref{iso}) will also be affected by the acceleration $A$. Consider, for instance, wave phenomena in a laboratory fixed on the Earth, which is usually assumed to follow a geodesic of the spacetime manifold. However, we wish to take into account a uniform nongravitational acceleration of the center of mass of the Earth. Indeed, the deviation of the Earth's motion from a geodesic could be due to the solar radiation pressure. Another possibility would be the Mathisson - Papapetrou coupling of the curvature of the solar gravitational field with the angular momentum of the Earth. Estimates suggest that such accelerations are very small and at a level below $\sim 10^{-10}$ cm/s$^2$. Nevertheless, for an Earth - based experiment the appropriate exterior field for the \lq\lq nonrotating spherical" Earth would be given by (\ref{iso}) if the acceleration can be considered uniform for the duration of the experiment. Taking the rotation of the Earth into account, we mention that the rotating C-metric has been discussed by a number of authors \cite{ES}, but is beyond the scope of this paper.

The Earth's acceleration will introduce a very small shift in the phase of a wave propagating in the gravitational field of the Earth. Consider, for instance, the gravitationally induced quantum interference of neutrons as in the COW experiment \cite{colella, RauchandWerner}. Let us imagine for the sake of simplicity that the $\hat Z$- axis of the system $\{ T, \hat X, \hat Y, \hat Z\}$ of metric (\ref{iso}) 
makes an angle $\Theta$ with the vertical direction in our local laboratory and so an otherwise free particle in the laboratory is subject to the effective Newtonian gravitational acceleration $g-A\cos \Theta$.
The corresponding neutron phase shift in the COW experiment would then be given by
\begin{equation}
\label{cownew}
\Delta \varphi =(g - A\cos\Theta) \frac{\mathcal{A}\omega}{v}\sin\alpha ,
\end{equation}
where $\omega$ is the de Broglie frequency of the neutron,  $\mathcal{A}$ is the area of the interferometer, $\alpha $ is the inclination angle of the interferometer plane with respect to the horizontal plane in the laboratory and $v$ is the neutron speed. When $A = 0$ or $\Theta=\pi/2$, this formula reduces to the standard formula of the COW experiment \cite{RauchandWerner}.

It should be mentioned that observational evidence already exists for the acceleration - induced phase shift of neutrons according to the experimental results of Bonse and Wroblewski \cite{BW} involving a system undergoing linear acceleration. The essential ideas in the derivation of (\ref{cownew}) can be naturally extended to other neutron experiments \cite{NES1,NES2} as well as interference experiments involving photons \cite{TAN} and atoms \cite{ATM1, ATM2,ATM3}. In particular, in future space - borne atom interferometry experiments the inertial acceleration of atoms resulting from the nongeodesic motion of the satellite should be taken into account along the lines indicated here.

Our discussion of neutron interferometry experiments has involved unpolarized neutron beams. In a noninertial frame of reference, the neutron is in general influenced by the translational and rotational accelerations of the system. The intrinsic spin of the neutron is expected to couple with the rotation of the frame. An interesting question is whether there exists a similar {\it direct} coupling between intrinsic spin and linear acceleration. At present, there is no observational evidence in support of a direct spin-acceleration coupling \cite{bm00}. 

\section{Pioneer anomaly}

The physical relevance of the C-metric may be clarified via nonrelativistic celestial mechanics as follows. Imagine an inertial reference frame and a star of mass $m$ such that its center of mass accelerates with a constant acceleration $\mathbf{A}=A \hat{\mathbf{z}}$ with $A>0$.
Thus the motion of a planet or a satellite about the star in terms of a noninertial coordinate system $\{t,x,y,z\}$ in which the star is at rest with its center of mass at the origin of the spatial coordinates is given to lowest order by
\begin{equation}
\frac{\rmd^2 \mathbf{r}}{\rmd t^2}+ \frac{m \mathbf{r}}{r^3}=-\mathbf{A}
\end{equation}
in accordance with Newtonian physics. 
The effective Newtonian gravitational potential is in this case given by $-m/r+Az$, which is analogous to the electric potential for the Stark effect.
Assuming that $-\mathbf{A}$ produces a small perturbation on the Keplerian motion of the planet or satellite, one can use the general methods of celestial mechanics, such as the scheme developed in \cite{bm}, to find the perturbed motion. Within the context of general relativity, the equation of motion of the test planet or satellite is given by the geodesic equation in the vacuum C-metric.

Let us now apply these ideas to the anomalous acceleration of Pioneer 10 and Pioneer 11 \cite{anderson1, anderson2}. 
The Pioneer 10/11 Missions were launched over thirty years ago and have been the first to explore the outer solar system. The analysis of Doppler tracking data from Pioneer 10/11 spacecraft is consistent with the existence of a small anomalous acceleration of about $10^{-7}\,$ cm/s${}^2$ toward the Sun. The effect first appeared at a distance of about 20 astronomical units from the Sun, where the outward acceleration of the spacecraft due to the solar radiation pressure reached a level well below $10^{-7}\,$ cm/s${}^2$.
To explain the Pioneer anomaly, the components of the acceleration $- \mathbf{A}$ along the directions of motion of the spacecraft must be both towards the Sun and approximately equal to $\sim 10^{-7}$ cm/s${}^2$. Let us recall that Pioneer 10 and Pioneer 11 are moving away from the solar system in almost opposite directions; more exactly, Pioneer 11 is out of the ecliptic, 17${}^\circ$ inclination, while Pioneer 10 is in the ecliptic, 3${}^\circ$ inclination \cite{d2}. 
Let $\hat{\mathbf{P}}$ and $\hat{\mathbf{P}}'$ be unit vectors that indicate the radial directions of motion of Pioneer 10 and Pioneer 11 with respect to the Sun, respectively. Suppose that the smaller angle between these directions is given by $\pi - 2 \beta$, where $\beta \simeq 7^\circ$. Then, $\mathbf{A}$ can be expressed as
\begin{equation}
\label{eq:56}
        \mathbf{A} = \frac{A_0}{2 \sin \beta }(\hat{\mathbf{P}}+\hat {\mathbf{P}}'), 
\end{equation}
where $A_0 \simeq 10^{-6}$ cm/s${}^2$ is the magnitude of the vector $\mathbf{A}$ and is such that, with $\sin 7^\circ \simeq 0.12$, $A_0 \sin \beta$ is the magnitude of the anomalous acceleration. Let us note using equation (\ref{eq:56}) that 
$\mathbf{A}\cdot \hat{\mathbf{P}} = \mathbf{A}\cdot \hat{\mathbf{P}}' = A_0 \sin \beta$.
It is therefore possible to find a vector $-\mathbf{A}$ that generates the Pioneer anomaly; however, the problem is then shifted to explaining the origin of such an acceleration of the center of mass of the Sun.

One possibility could be recoil acceleration due to the anisotropic emission of solar radiation.
For most practical purposes, one may assume that normal stars on the main sequence such as the Sun radiate isotropically. As a matter of principle, however, it is difficult to believe that at any given instant of time the net momentum radiated along all antipodal directions would be exactly zero. 
However, even if all of the Sun's intrinsic luminosity were directed only along the negative $z$-axis, the recoil of the Sun due to momentum conservation would have an acceleration $A\hat{\mathbf{z}}$, where $A\simeq  10^{-10}$ cm/s${}^2$. 
This is about four orders of magnitude smaller than the acceleration needed to explain the Pioneer anomaly.
On the other hand, the Sun also emits charged particles in the form of the solar wind, coronal mass ejections, etc.; again, simple estimates suggest that the Pioneer anomaly is comparatively too large to be explained by any anisotropy in normal solar activity. 
Thus it appears highly unlikely on the basis of current data that solar recoil acceleration could be responsible for the Pioneer anomaly.
In any case, the C-metric is in principle no longer applicable to a radiating source.
These issues require further investigation.

\section{Discussion}

In this paper, we have considered the possibility that the Pioneer anomaly could be due to a small uniform acceleration of the center of mass of the Sun. Within the framework of general relativity, accelerating gravitational sources have been discussed by many authors \cite{ES}. The simplest such solution corresponding to a uniformly accelerated Schwarzschild source  is the C-metric. We have shown that the behavior of particles and waves on the linearized C-metric background can be described by the gravitational Stark effect.

\section*{Acknowledgements}
The authors acknowledge R. Ruffini for his encouragement and support. BM is grateful to John D. Anderson for discussions concerning the Pioneer anomaly. CC wishes to thank J. Griffiths and G. Scarpetta for useful discussions.

\end{document}